
\input phyzzx
\def\bco#1{{2#1\choose #1}}
\def\rfour#1{\left({R\over4}\right)^{#1}}
\def\gtil#1{\tilde g_{#1}}
\hyphenation{cor-rec-tions}

\Pubnum={\vbox{ \hbox{CERN-TH-6600/92}\hbox{FTUAM-9223}}}
\date={July, 1992}
\pubtype={}
\titlepage

\title{A PROPOSAL FOR STRINGS AT $D>1$}
\vskip 1.0cm
\author{L. Alvarez-Gaum\'e \break J.L.F. Barb\'on
\foot{ Permanent address: Departamento
de F\'{\i}sica Te\'orica,
Universidad Aut\'onoma de Madrid,
Canto Blanco, Madrid. Spain. } \break \v C. Crnkovi\'c \break}

\address{Theory Division, CERN\break
 CH-1211 Geneva 23, Switzerland}

\abstract{ Using the reduced formulation
of large-$N$ Quantum Field Theories we study strings in
space-time dimensions higher than one.	We present results
on possible string susceptibilities, macroscopic loop operators,
$1/N^2$-corrections and other general
properties of the model.}

\endpage

\pagenumber=1

\chapter{INTRODUCTION AND CONCLUSIONS}

One of the most interesting open problems in String Theory and
Quantum Field Theory is the study of the systems with string
excitations in dimensions below the critical dimension.
In the context of lattice gauge theories the problem
can be formulated in terms of the Schwinger-Dyson
equations satisfied by the Wilson-loop observables.
In the context of large-$N$ field theory, it was noticed
in the early eighties that due to the factorization
properties of most gauge invariant observables,
the loop-equations of the theory were equivalent
to those of a theory on a single hypercube
with periodic boundary conditions.  These models are
known as the
large-$N$ reduced models \REF\ek{T. Eguchi and H. Kawai,
Phys. Rev. Lett.{\bf 48} (1982) 1063.}
\REF\tek{A. Gonzalez-Arroyo and M. Okawa, Phys. Rev. {\bf D 27}
(1983) 2397, Phys. Lett. {\bf B 133}(1983) 415, Nucl. Phys.
{\bf B 247} (1984) 104.}
\REF\das{ For a review with references to the literature see
S.R. Das, Rev. Mod. Phys. {\bf 59} (1987) 235.}
[\ek,\tek,\das].

In this paper we use the techniques of the reduced models
to study non-critical
strings in dimensions higher than one.	We will find an
action in terms of a single matrix which exactly reproduces
the partition function of a string moving in $2,3,4$ space-time
dimensions.  The model does not appear yet to be exactly
solvable, and we can at the moment make only general
preliminary remarks concerning its critical properties
and its possible continuum limit.

The first lattice formulations of string theories
in arbitrary dimensions appeared
in the mid-eigthies
\def\npb{Nucl. Phys.\ }
\def\prd{Phys. Rev.\ }
\def\prl{Phys. Rev. Lett.\ }
\def\plb{Phys. Lett.\ }
\REF\adf{J. Ambj\o rn, B. Durhuus and J. Fr\"ohlich,
\npb {\bf B 259} (1985) 433.}
\REF\adfj{J. Ambj\o rn, B. Durhuus and J. Fr\"ohlich
and P. Orland, \npb {\bf B 270} (1986) 457.}
\REF\david{F. David, \npb {\bf B 257} (1985) 53,
\npb {\bf B 257} (1985) 543.}
\REF\kazakov{V.A. Kazakov, \plb {\bf 150 B} (1985) 282.}
\REF\kpz{A.M. Polyakov, Mod. Phys. Lett. {\bf A2} (1987) 893;
V. G. Knizhnik, A.M. Polyakov and A.B. Zamolodchikov,
Mod. Phys. Lett. {\bf A 3} (1988) 819.}
\REF\ddk{F. David, Mod. Phys. Lett. {\bf A 3} (1988) 1651;
J. Distler and H. Kawai, \npb {\bf B 321} (1988) 509.}
[\adf,\adfj,\david,\kazakov], and there was a good deal of
activity in solving some two-dimensional models
on random triangulated surfaces
\REF\models{V.A. Kazakov, \plb
{\bf 119 B} (1986) 140, D. Boulatov
and V. A. Kazakov, \plb{\bf 186 B} (1987) 379;
I.K. Kostov, and M.L. Mehta, \plb {\bf 189 B} (1987) 118;
B. Duplantier and I. Kostov, \prl {\bf 61} (1988) 1433; etc.}
[\models].  A breakthrough took place with the work in
ref. [\kpz] which provided a description in the continuum
of the coupling of minimal conformal models to two-dimensional
gravity in the light-cone gauge.  Their subsequent formulation
in the conformal gauge, and the generalization
of the results to surfaces of arbitrary topology appeared
in [\ddk]. With the discovery of the double scaling limit
\REF\doubles{M.R. Douglas, S. Shenker,
\npb {\bf B 335} (1990) 635;
E. Br\'ezin and V.A. Kazakov, \plb {\bf 236 B} (1990) 144;
D. Gross and A.A. Migdal, \prl {\bf 64} (1990) 127,
\npb {\bf B 340} (1990) 333.}
[\doubles] we begun to understand some of the perturbative
and non-pertubative properties of non-critical strings
in dimensions below or equal to one.  Little progress has been
made however in going beyond $d=1$ apart from some evidence
that it is possible to construct Liouville Quantum Field
Theory in some special dimensions $d=7,13,19$
\REF\gervais{See J.-L. Gervais, Comm. Math. Phys. {\bf 138}
(1991) 301; Phys. Lett. {\bf 243 B} (1990) 85,
and references therein.}.  Some interesting recent
results appear in
\REF\brezin{E. Br\'ezin and S. Hikami, {\it A Naive Matrix-Model
Approach to Two-Dimensional Quantum Gravity Coupled to Matter of
Arbitrary Central Charge}, LPTENS preprint 92/10.}
[\brezin].

Some interesting problems involving the presence of external field
have also been studied
\REF\grossbre{E. Brezin and D. Gross, \plb {\bf  97 B}
(1980) 120.}
\REF\grossnew{D. Gross and Newman,
\plb {\bf 266 B} (1991) 291;
PUPT-1282, Dec. 91.}
\REF\wittenkon{E. Witten, Surveys in
Diff. Geom. {\bf 1} (1991)
 243, {\it On the Kontsevich Model and Other Models of
Two Dimensional Gravity}.  IAS preprint HEP-91/24.  M.
Kontsevich,
Funct. anal. i ego pril. {\bf 25} (1991) 50;
{\it Intersection
Theory on the Moduli Space of Curves and the Matrix
Airy Function}.  M. Plank Institute
for Mathematics Preprint,	October 91.}
\REF\tata{S.R. Das, A. Dhar, A.M. Sengupta and S.R. Wadia,
Mod. Phys. Lett. {\bf A5} (1990) 1041.}
[\grossbre,\grossnew,\wittenkon,\tata].  In particular
the work in [\tata] attempted to generalize the standard
one-matrix model to include the presence of local curvatures
in the triangulated surfaces described by the
 large-$N$ limit.  We will
make contact with this work later in sections three and four.

The outline of this paper is as follows: In section two
we formulate the problem of strings in arbitrary
dimensions in general, we present the general methodology
of reduced actions, and the explicit form of the one-matrix
model action which reproduces exactly the properties of
planar string in $D$-dimensions.  Some details
are given concerning the way to carry out perturbative
computations in the context of reduced models.

In section three we continue the analysis of our model and show
how in the simplest possible approximation one obtains
a second order phase transition between a pure
gravity phase with string susceptibility
$\gamma_{{\rm st}}=-1/2$
and a branched polymer phase with
$\gamma_{{\rm st}}=+1/2$.  This is in agreement
with the analysis carried out in [\tata] for the
one-matrix model with world-sheet curvature.

In section four we analyze some general
effective actions which could be taken as
approximations to our theory.  We solve the
planar limit of these theories exactly by
looking for one-cut solutions.	We find
generically a second order phase transition between a
pure gravity-like phase with negative $\gamma_{{\rm st}}$
(with the value $\gamma_{{\rm st}}=-1/(m+1)$, as in
the Kazakov multicritical points [\kazakov],
and a phase that resembles branched polymers with
positive string susceptibility $\gamma_{{\rm st}}=n/(n+m+1)$,
with $n,m$
arbitrary positive integers.  In the approximation we envisage,
the $D$-dimensional embedding of the string is effectively
replaced by some kind of polymeric matter whose role is
to break the world-sheet into surfaces oscullating
at several points.  We have the following phases: a phase where
gravity becomes
critical and matter remains away from criticality, in which
case we obtain the pure gravity regime; a phase where
matter becomes critical while gravity remains non-critical, leading
to a theory of branched polymers; and finally the most
interesting phase where both become critical simultaneously.
This phase is described by the positive string susceptibility
$\gamma_{{\rm st}}=n/(n+m+1)$.

To understand better the geometrical properties of this phase, we
compute correlators of macroscopic loop operators on the surface in
section five. We show that it is possible to define macroscopic loops
in the continuum limit only for a subclass of critical points with
$n=1$. We find it quite remarkable that only for the $n=1$ critical
points is there enough room on the surface to open arbitrary numbers
of loops of arbitrary macroscopic length.  For higher values of $n$,
the polymerization of the surface is presumably so complete that no
room is left to construct such operators. This leads us to believe
that only in this phase we may be able to recover some of the basic
properties of non-critical strings.

The continuum limit for $n=1$
contains an extra state with respect to the standard pure gravity
case, representing the breaking of the surface into two pieces
touching at one point. This  extra state is probably the shadow of the
effect of the tachyon that one expect to find as soon as string
theories are formulated in dimensions higher than one. Whether this
feature remains when more realistic effective actions are considered
remains to be seen, but we find it rather encouraging. We
also explore some of the scaling operators appearing in truncations of
the reduced model.

To make sure that the scaling limit we take indeed defines a
string theory, in section six we solve the Schwinger-Dyson equations
for the approximate effective action on genus
one.  We find that the David-Distler-Kawai
dependence of the string susceptibility on the
genus of the surface is recovered to this order.
This is encouraging evidence that the continuum
limit that we define does describe some kind of non-critical
strings.

Although we find the results just summarized encouraging, there are
several obvious points in our approximations which should be
clarified.  First of all, in the effective actions studied in section
four there is little evidence of the dependence of the string
susceptibility on the string embedding dimensions.  In particular, in
the reduced model one of the couplings represents the two-dimensional
cosmological constant.	There is a similar coupling in the effective
action for the reduced model which plays a similar role.  What needs
to be done is to show that the relation between these two couplings is
analytic, at least for some range of parameters.  The conclusions we
draw in sections four and five depend implicitly on this assumption,
and this is an important caveat to keep in mind when considering the
work presented.  The most natural thing to do seems to be to try to
settle these questions by a direct renormalization group analysis of
the reduced model along similar lines to those suggested in
\REF\zinbre{E. Brezin and J. Zinn-Justin, {\it Renormalization Group
Approach to Matrix Models}, preprint LPTENS 92/19,
SPhT/92-064.}[\zinbre], or to construct an approximate
Migdal-Kadanoff-type transformation.  Work in this direction is in
progress. We hope to convince the reader that the study of
non-critical strings in dimensions higher than one, using reduced
large-$N$ model is worth pursuing.

While this work was being completed we received several preprints
having some overlap with our work \REF\korchem{G. P. Korchemsky, {\it
Matrix Model Perturbed by Higher Order Curvature Terms}, Parma
preprint UPRF-92-334, {\it Loops in the Curvature Model}, preprint
UPRF-92-340.}[\korchem].

\chapter{LARGE-$N$ REDUCED THEORIES AND MATRIX MODELS}

In the study of strings propagating in flat $D$-dimensional
space-time, we represent the sum over two-dimensional metrics
in Polyakov's approach to string theory
\REF\polyakov{A.M.Polyakov,\plb {\bf 103 B} (1981) 207, 211. }
[\polyakov] by a sum over
tringulated surfaces.	The quantity we would like to evaluate
[\adf,\david,\kazakov] is
$$
Z=\sum_g \kappa^{2(g-1)} \sum_T {e^{-\mu |T|}\over |n(T)|}
\int \prod_{i\in T_0}\sigma_i^{\alpha} d^DX_i
\prod_{\langle ij\rangle\in T_1} G(X_i-X_j)
\eqn\stringsum
$$
Where $g$ is the genus of the triangulation.  For a given
triangulation $T$, $T_0,T_1,T_2$ are respectively
the sets of vertices, edges and faces of $T$, $n(T)$ is the
order of the symmetry group of $T$, $|T|$ is the
total area of $T$ counting that every triangle has unit
area.  $\mu$ is the bare
two-dimensional cosmological constant,
$X_i$ describes the embedding of the triangulation into
$D$-dimensional flat space-time and  $G(X-Y)$ is the
propagator factor for each link.  We have included also the
local volume factor $\sigma_i ^{\alpha}$ at site $i$ to
represent the effect of local curvature. Usually we take
$\alpha=D/2$, but it is more convenient
to leave this exponent arbitrary
to include the effect of local curvature terms on the
world-sheet. For a triangulation, if $q_i$ is the local
coordination number at site $i$ (the number of triangles
sharing this vertex),
$$
\sigma_i={q_i\over 3},
$$
and the volume factor in the measure can be shown to be related
in the continuum to an expansion of the form:
$$
\sum_i \ln \sigma_i =c_0 +c_1\int \sqrt{g} R + c_2\int\sqrt{g}
R^2\ldots ,
$$
thus representing some of the effects of world-sheet curvature.  The
general features of the phase diagram in the $(D,\alpha)$ plane
were studied for example in
\REF\bkkm{D. Boulatov, V.A. Kazakov, I. Kostov, A.A. Migdal,
\npb {\bf B 275} (1986) 641.}
[\bkkm].  In the standard Polyakov formulation, the propagator
is Gaussian,
$$
G(X)=e^{-X^2/2}
$$
By standard large-N analysis the sum \stringsum\  can be transformed
[\david,\kazakov] into the large-$N$ expansion of a matrix
field  theory in $D$-dimensions
$$
Z=\lim_{N\rightarrow \infty} \log\int D\phi
\exp \left(-N\int d^D X Tr({1\over 2}\phi G^{-1}\phi
+{1\over 3} g\phi^3)\right)
\eqn\mastersum
$$
It is difficult to proceed very far with Gaussian propagators.	If
we work in dimensions  $D<6$ it should not matter whether we
replace the Gaussian propagator by the Feynman propagator.
There is numerical evidence that this change does not affect
the critical properties of the theory below six dimensions
\REF\boulkaz{D. Boulatov, V.A. Kazakov,
\plb {\bf 214 B} (1988) 581;
J. Ambj\o rn, Acta Phys. Pol. {\bf B21} (1990) 101,
and references therein.}
[\boulkaz].  Notice that in this construction the exponent
$\alpha$ in \stringsum\ is set to zero.  If we want to include the
effect of local world-sheet curvature, we can follow [\tata]
and change the kinetic term in \mastersum.  The two-dimensional
cosmological constant is here represented by $g$, $g=e^{-\mu}$.  To
summarize, we want to study the critical properties of
the action:
$$
Z=\int D\phi(X) \exp \left(-N\int d^D X Tr({1\over
2}\partial_{\mu}\phi \partial^{\mu}\phi+{g\over 3}\phi^3)\right)\ \ ,
\eqn\masterone
$$
or
$$
Z=\int D\phi(X) \exp \left(-N\int d^D X Tr({1\over 2}
A \partial_{\mu}\phi
 A\partial^{\mu}\phi+{g\over 3}\phi^3)\right)\ \ ,
\eqn\mastertwo
$$
where $A$ is a constant $N\times N$ matrix and $\phi$ is an
$N\times N$ matrix field.  The reason why the $A$-matrix in the
kinetic term simulates the effect of local curvature can be
seen by writing the propagator in \mastertwo\ explicitly
$$
\langle \phi_{ij}(X) \phi_{kl}(Y)\rangle = A^{-1}_{jk}A^{-1}_{li}
G(X-Y).
$$
If we ignore for the time being the propagator factor $G$, for
every closed index loop in a generic $\phi^3$ graph made of
$q$ propagators, we obtain a contribution of $TrA^{-q}$.
Since the $\phi^3$ graphs are dual to triangulation, this
means that we are associating a curvature factor of
$TrA^{-q}$ to the vertex dual to the face considered.  In this
way we can simulate the presence of the $\sigma^{\alpha}$ term
in the measure in \mastersum. We are interested in particular
in the computation of the string susceptibility exponent
$$
\chi \sim {\partial^2F\over\partial g^2}
\sim (g-g_c)^{-\gamma_{\rm st}} ,
\eqn\stringsus
$$
where $F$ is the free energy of the system, and $g_c$ is the
critical value of $g$ indicating the location of the critical
point.	Apart from the string susceptibility we would also like
to compute the spectrum of scaling operators and the properties
of the quantum geometry implied by \masterone \mastertwo.

To simplify the arguments we concentrate on the study of
planar configurations (spherical topologies).  From work dating
back to the late seventies
\REF\witten{E. Witten, {\it Recent Developments
in Gauge Theories},
Proceedings of the 1979 Carg\`ese
Summer Institute.  G. 'tHooft et al. eds.
Plenum Publ. New York.}[\witten], and due
to the factorization properties of the leading large-$N$
approximation, the planar limit is dominated by a single constant
configuration; Witten's master field (a master orbit in the case
of gauge theories).  This idea is explicitly realized in
$0$-dimensional matrix models
\REF\brezin{E. Br\'ezin C. Itzykson, G. Parisi
and J.B. Zuber, Comm. Math. Phys. {\bf 59} (1979) 35.
} [\brezin]. In the
lattice gauge  theories one can think of the reduced Eguchi-Kawai
(EK)[\ek] or Twisted Eguchi-Kawai (TEK)[\tek] models as explicit
descriptions of the master field.  We now briefly describe
the main ideas behind the reduced models.  Further details
can be found in the previous
references and in the review article
[\das].  These models
were originally formulated to describe the large-$N$
properties of lattice
gauge theories.

Due to the factorization property in the
large-$N$ limit, the Schwinger-Dyson
equations for Wilson-loops form a closed system of equations
in the planar limit.  For gauge theories we get an infinite set
of polynomial equations for the Wilson-loops
\REF\loopeq{Yu. M. Makeenko and A.A. Migdal,
\plb {\bf 88 B} (1979) 135; S. Wadia,\prd {\bf D 24} 1981) 970.  For
more details and references see
A.A. Migdal, Phys. Rep. {\bf 102} (1983) 200.}
[\loopeq].  In terms of the standard link variables
$U_{\mu}(x)$, the EK proposal consists of making them
independent of $x$, $U_{\mu}(x)\mapsto U_{\mu}$, and the
action of the reduced model becomes
$$
S_{EK}={1\over {\rm Vol.}}S(U_{\mu}(x)\mapsto U_{\mu}).
\eqn\sek
$$
Some of the properties of this action are:

1).  We obtain a theory on a single hypercube with periodic boundary
conditions.  This is an important simplification of the problem.

2).  In the case of gauge theories, the gauge symmetry becomes a global
symmetry, $U_{\mu}\mapsto \Omega U_{\mu} \Omega^{-1}$.

3).  There is an extra $U(1)^D$ symmetry in the case
of $U(N)$ lattice gauge theory $U_{\mu}\mapsto e^{i\theta_{\mu}}
U_{\mu}$.  In the case of $SU(N)$ the symmetry becomes
$Z_N^D$.

4).  It is possible to show that the planar
Schwinger-Dyson equations
following from the reduced action coincide with those obtained
from the original Wilson theory as long as open loop expectation
values vanish.	This can be shown to be true at strong coupling,
but it does not hold at weak coupling
\REF\heller{ G. Bhanot, U. Heller, H.Neuberger, \plb {\bf 113 B}
(1982) 47.} [\heller].	Without this
problem we would have a beautiful
implementation of the Master Field idea, because the
loop equations for the 1-site EK model are identical with
the standard ones.  The problem is connected with the fact
that at weak coupling the extra $U(1)^D$ symmetry is broken.
Open loops have non-trivial charge with respect to this
symmetry.  Only if the symmetry
remains unbroken are we
guaranteed to maintain the loop equations without extraneous
terms.

The resolution of this problem motivated the
formulation of the TEK model [\tek].  It is inspired on
'tHooft's use of twisted boundary conditions in gauge
theories
\REF\thooft{G. 'tHooft, \npb {\bf B 153} (1979) 141, Acta Phys.
Austriaca, Suppl. {\bf 22} (1980) 531.}
[\thooft].  For any matrix theory the TEK prescription
is exceedingly simple.	We reduce according to
$$
\phi(x)\mapsto D(x) \phi D(x)^{-1},
\eqn\tekred
$$
where $D(x)$ is a projective representation of the
$D$-dimensional lattice translation group.  We have
to choose a set of $D$ $N$-dimensional matrices
$\Gamma_{\mu}$ such that:
$$
D(x)=\prod_{\mu}\Gamma_{\mu}^{x_{\mu}},\qquad
x=(x_1,x_2,\ldots,x_D), \qquad x_i\in Z,
\eqn\dtrans
$$
since $D(x)$ has an adjoint action on the fields $\phi$,
the $\Gamma_{\mu}$'s are required to commute
only up to an element of the center of $SU(N)$,
$$
\Gamma_{\mu}\Gamma_{\nu}=Z_{\nu\mu}\Gamma_{\nu}\Gamma_{\mu},
$$
$$
Z_{\mu\nu}=e^{2\pi i n_{\mu\nu}/N},
\eqn\tmatrices
$$
and the integers $n_{\mu\nu}$ are defined mod $N$.  The reduced
action prescription is now
$$
S_{TEK}={1\over {\rm Vol.}}S(\phi(x)\mapsto D(x)\phi
D(x)^{-1}),
\eqn\tekaction
$$
and similarly for expectation values.

To avoid open loop expectation
values and a mismatch between the original and reduced
Schwinger-Dyson equations, the matrices $\Gamma_{\mu}$
must verify some conditions.
In particular they should
generate an irreducible representation
of the group of lattice translations.
For a lattice with $L^D$ sites this
requires $N=L^{D/2}$.  Hence
$N=L$ for $D=2$ and $N=L^2$ for $D=4$.	Hence, if we want to
simulate a two-dimensional lattice with $64\times 64$ sites
it suffices to consider the group $SU(64)$.  The choice
of twist matrix $n_{\mu\nu}$ depends on the dimensionality.
In two-dimensions, the simplest
choice is given by
$$
n_{\mu\nu}=\epsilon_{\mu\nu}.
\eqn\twodtwist
$$
The explicit form of the four and higher dimensional twists
can be found in the quoted literature.

We now set aside lattice gauge theory and return to our problem.
The twisted reduced version of \masterone\ becomes:
$$
S={1\over 2}\sum_{\mu}Tr(\Gamma_{\mu}\phi\Gamma_{\mu}^{-1}
-\phi)^2+TrV(\phi).
\eqn\rstring
$$
The equivalence of the planar approximation to \masterone\ with
\rstring\ follows from earlier work on reduced models
\REF\en{T. Eguchi and R. Nakayama, \plb {\bf 122 B} (1983) 59.}
[\tek,\en].  For simplicity we consider the two-dimensional
case.  First, since for $k_{\mu}=\epsilon_{\mu\nu}q_{\nu}$
and $q_\nu$ defined modulo $N$,
$$
A(q)=\Gamma_1^{k_1}\Gamma_2^{k_2}
$$
are $N^2$ linearly independent matrices,   we can expand the
matrix $\phi$ in the $\Gamma$-basis according to
$$
\phi=\sum\phi_q A(q).
$$
Some useful properties of the $A$-matrices are:
$$
A(q)^{\dagger}=A(-q)e^{{2\pi i \over N}}\langle k|k\rangle,
\qquad \langle k^i|k^j\rangle =\sum_{\mu<\nu}n_{\mu\nu}
k_{\nu}^ik_{\mu}^j = k^i_2 k^j_1,
$$
$$
A(q^1,\ldots, q^n)=A(q^1 + \cdots + q^n)
\exp \left({{2\pi i\over N}}
\sum_{i<j} \langle k^i | k^j\rangle\right),
$$
 and
$$
\Gamma_{\mu}A(q)\Gamma_{\mu}^{\dagger}=e^{2\pi i q_{\mu}/L}A(q) ,
$$
explicitly showing how the $\Gamma$'s implement the lattice
translation group.  Furthermore,
$$
Tr(A(q)^{\dagger}A(q')) = N\delta_L(q-q'),
$$
$$
Tr(A(q^1)\ldots A(q^n)) = N \delta(\sum q^i)
\exp \left({{2\pi i\over N}}
\sum_{i<j} \langle k^i | k^j\rangle\right),
$$
where the $\delta$-function is defined mod $L$.  Using these
properties, the kinetic term
of our action becomes
$$
{1\over 2}\sum_{\mu}Tr(\Gamma_{\mu}\phi\Gamma_{\mu}^{-1}
-\phi)^2 +{1\over 2}m^2Tr\phi^2\sim
$$
$$
N\sum_q\left[ {1\over 2}m^2 +\sum_{\mu}(1-\cos ({2\pi q_{\mu}
\over L})\right] \phi_q \phi_{-q} e^{-2\pi i \langle k|
k\rangle /N},
\eqn\prop
$$
coinciding with the standard lattice propagator of
a scalar field on a size $L$ $D$-dimensional lattice
up to the last phase factor in \prop.

The role of this
and similar phase factors appearing in the vertices
of the theory is to restore the topological expansion
in large-$N$.  We have now obtained an effective
field theory with scalar variables $\phi_q$.
The phase factors
keep track of the fact that the theory came from
a matrix field theory and that the expansion
in powers of $1/N$ can be represented in terms
of a topological expansion where the leading order
corresponds to spherical topologies, and higher orders
come from surfaces of increasing Euler number.

In particular, the  phases of all planar graphs vanish. The easiest
way to see that is to note that the phase factor does not change when
we contract all the propagators connecting two vertices. The two
vertices  contribute factors $Tr[A(q_1)\ldots A(q_n)$$A(q'_1)\ldots
A(q'_{n'})]$ and $Tr[A(-q'_1)\ldots A(-q'_{n'})$$A(q''_1)\ldots
A(q''_{n''})]$ while the phase factor coming from the propagators is
$\exp \left({{2\pi i\over N}} \sum_{i=1}^{n'} \langle k'_i |
k'_i\rangle\right)$. All together, the phase factor is
$$
\eqalign{
\delta(\sum_i q_i +\sum_i q'_i)\delta(-\sum_i q'_i+\sum_i q''_i)
\exp \biggl[{{2\pi i\over N}}\biggl(
\sum_{i<j} \langle k_i | k_j\rangle +\sum_{i,j}
\langle k_i | k'_j\rangle \cr
+\sum_{i<j}\langle k'_i | k'_j\rangle +\sum_i\langle k'_i |
k'_i\rangle -\sum_{i,j}\langle k'_i | k''_j\rangle +\sum_{i<j}
\langle k''_i | k''_j\rangle +
\sum_{i<j} \langle k'_i | k'_j\rangle \biggr)\biggr]\cr
\propto\exp \biggl[{{2\pi i\over N}}\biggl(
\sum_{i<j} \langle k_i | k_j\rangle +\sum_{i,j}
\langle k_i | k''_j\rangle +
\sum_{i<j}\langle k''_i | k''_j\rangle \biggr)\biggr],\cr}
$$
where we used the $\delta$-function constraints. This is precisely the
phase factor of a single vertex with $\{k_i\}$,  $\{k''_j\}$
propagators attached. Provided we are dealing with a planar graph, we
can repeat this procedure until we end up with a single vertex with no
propagators attached whose phase factor is trivially zero.

In case of non-planar graphs we will not be able to contract the
propagators winding around non-trivial homology cycles. For example,
it is easy to see that the phase corresponding to the torus diagram
consisting of two propagators with momenta $q_1$ and $q_2$, each
following one of the two non-trivial homology cycles and meeting in
one vertex, is
$$
\exp\left[{2\pi i\over N}\left(\langle k_1|k_2\rangle -
\langle k_2|k_1\rangle \right)\right].
$$
The non-vanishing phases of non-planar diagrams result in their
$1/N^2$-suppression. Finally, note that \prop\ shows that
 the space-time degrees
of freedom of the original theory are coded in the
Fourier transforms of the ``internal" $SU(N)$ indices.

The construction we have just carried out works only
for even dimensions.  For odd-dimensions we can
leave one of the dimensions unreduced, and apply
the reduction prescription to the remaining
even number of dimensions.  This yields an action
\def\onet{{1\over 2}}
$$
S=\int dt \onet Tr\dot\phi(t)^2+\onet\int dt Tr
(\Gamma_{\mu}\phi(t)\Gamma_{\mu}^{-1}
-\phi)^2 +\int dt Tr V(\phi(t)).
$$
We can consider a slightly more general model by
including a general coupling in the hopping term:
$$
-S(\phi)=\sum_{\mu}\onet R^2 Tr\phi\Gamma_{\mu}\phi\Gamma_{\mu}
^{\dagger}-\onet m^2\phi^2-TrV_0(\phi),
$$
$$
V_0(\phi)={g_3\over \sqrt{N}}\phi^3+{g_4\over N}\phi^4+\ldots,
\qquad m^2=2DR^2+m_0^2.
\eqn\raction
$$
The hopping term is the first term in $S(\phi)$.  Adding the
effect of world-sheet curvature is easy, we simply
replace the hopping term according to
$$
Tr(\Gamma_{\mu}\phi\Gamma_{\mu}
^{\dagger}-\phi)^2\mapsto
TrA (\Gamma_{\mu}\phi\Gamma_{\mu}^{\dagger}-\phi)
A(\Gamma_{\mu}\phi\Gamma_{\mu}^{\dagger}-\phi)
\eqn\ractioncur
$$
Our proposal to describe strings in dimensions higher than
one is to investigate the properties of \raction\ and
\ractioncur. We have effectively reduced the problem
to a one-matrix model.	The
difficulty lies however in the presence of the twist
matrices.  We begin our analysis in the following section.

\chapter{PROPERTIES OF THE REDUCED STRING ACTION}

Some general features of \raction\ and
\ractioncur\ are easy to
extract.  If the hopping term is
very small, we are in the limit
where the space-time lattice points are
very far apart and the
whole surface is mapped into a single point,
or various points
incoherently.  This is the pure gravity
regime, $\gamma_{\rm st}$
\def\gst{\gamma_{\rm st}}
is expected to be $-1/2$ and the induced metric plays no role.
As the hopping term is increased, it becomes more likely for
the surface to occupy more and more space-time lattice sites,
the induced metric and singular embeddings in principle
begin to play important roles, and we can expect a transition
to a phase different from pure gravity at some critical
value of the hopping coupling $R^2$.  We will show presently
that in the most na\"{\i}ve approximation we are driven
to a branched polymer phase with $\gst=1/2$,
and at the transition point $\gst=1/3$.  A more careful
analysis will show a much broader spectrum of possibilities.
Note that the role played by $R$ is similar to the compactification
radius in the study of $c=1$ strings propagating in a
circle of radius $R$
\REF\grosskleb{D. Gross and I. Klebanov, \npb {\bf B 344}
(1990) 475.}
[\grosskleb].

To deal with \raction\ we decompose the matrix $\phi$ into
its eigenvalues and angular variables as usual
$$
\phi=U^{-1}\lambda U\qquad \lambda=
{\rm diag}(\lambda_1,\lambda_2,\dots,\lambda_N),
$$
The angular integration produces an effective action
$$
e^{\Gamma_{\rm eff}(\phi)}=\int dU e^{R^2\sum_{\mu}Tr\lambda
\Gamma_{\mu}(U) \lambda \Gamma^{\dagger}(U)}
$$
$$
\Gamma_{\mu}(U)=U\Gamma_{\mu}U^{-1}
$$
$$
Z=\int d\phi e^{\Gamma_{\rm eff}[\phi]-V_0(\phi)}
\eqn\effaction
$$
The simplest representation of $\Gamma_{\rm eff}$ is given
in terms of the leading order expansion in $R^2$:
$$
Z=\int d\phi e^{N\left(R^2{D\over N}(Tr\phi)^2-V_0
(\phi)\right)}.
\eqn\polyaction
$$
The critical properties of \polyaction\ are very similar
to those of the curvature model in [\tata].  What the
first term in $\Gamma_{\rm eff}$ is describing is the
breaking of the surface.  The Feynman graph representation
of a term like $(Tr \phi^n)^2$ is a vertex where two
surfaces osculate at one point, and the coordination
number of the touching
vertices in each of the surfaces is $n$.
If we have instead a term of the form
$(Tr\phi^n)^p$, it represents $p$-surfaces sharing one
point.	When there are no products of traces, the Feynman
graph expansion for the free energy or the connected
correlators involves only the sum over connected orientable
surfaces.  When products of traces are included we
obtain an expansion where we have surfaces of different
topologies touching at points determined by the new
vertices.  The new graphs look like ``daisies".  In
the original curvature model introduced in [\tata]
the initial partition function is
$$
Z=\int d\phi e^{-N\left( \onet \phi A\phi A+g Tr\phi^4
\right)}
\eqn\curmodel
$$
The angle integral is again very difficult to carry out in
this model, and the simplest non-trivial approximation
is to study the action:
$$
S_{\rm eff}=N\left(\onet Tr\phi^2 +gTr\phi^4
+{g'\over N}(Tr\phi^2)^2\right).
$$
For real $A$, the coupling $g'=(TrA^2/N-(TrA)^2/N^2)$
is always positive.  In this case as we will see later
one cannot escape the pure gravity phase.  To leading
order in $1/N$, the graphs have the topology of
spheres touching at one point, where any two spheres
in the graph can share at most one common point, and
no closed loop of spheres can appear.  A generic
graph looks like a tree of spheres.
In the planar approximation it was
shown in [\tata] that we have the following possibilities:

1). $g'>0$, $\gst =-1/2$ and we are
in the pure gravity regime.

2). $g_c<g'<0$.  Again $\gst =-1/2$ and we continue
in the pure gravity case.  The branching is still
unimportant, and the leading contributions to
the free energy come from graphs with few big spheres
and few small branchings.

3).  At $g'=g_c=-9/256$, $\gst=+1/3$ and the number
of branchings become competitive with the number of
smooth spheres.

4).  For $g<g_c$ we obtain $\gst=+1/2$ and the dominating
term is the one with a product of two traces, thus
the notion of a smooth surface disappears and we end
up with a branch polymer phase.

The same analysis can be carried out with our na\"{\i}ve
approximation \polyaction, and we will only note
the differences with respect to the analysis in
[\tata] we have just reviewed.	The basic differences
are:

1).  Our potential is cubic, $V_0=\onet \phi^2+g\phi^3
/\sqrt{N}$, the one-cut solution is asymmetrical, and
the new vertex is just given by $(Tr \phi)^2$.	These
are purely technical differences and nothing in the
analysis changes substantially.

2).  We are always in the $g'<0$ region.  In the
original model in [\tata] we can reach this region
only after analytically continuing the matrix
$A\mapsto iA$ in order to obtain the polymer phase.
However, in this case the interpretation of the matrix
$A$ in terms of local world-sheet curvature loses its
meaning.

Our simplified
model then contains a pure gravity phase, an
intermediate phase with $\gst=1/3$, and a
branched polymer phase with $\gst=1/2$.  We should
like to remark that the exact solution to the
pure curvature model \curmodel\ is known only for
Penner's potential
\REF\makeenko{Yu. Makeenko and L. Chekhov,
{\it The Multicritical Kontsevich-Penner Model}.
NBI-HE-92-03, January 1992.}
[\makeenko], however in this case the full theory
has the symmetry $A\mapsto {\rm const.}A^{-1}$
making the interpretation of
$A$ in terms of world-sheet curvature rather
doubtful.

\chapter{PRELIMINARY ANALYSIS OF THE EFFECTIVE ACTION}

We have learned that the basic problem
is the evaluation of the angular integral
$$
e^{\Gamma_{\rm eff}}=
\int dU e^{R^2\sum_{\mu} Tr \lambda U\Gamma_{\mu}U^{-1}
\lambda U\Gamma_{\mu}^{\dagger}U^{-1}}.
\eqn\basicint
$$
The na\"{\i}ve approximation embodied in \polyaction\
already gave us a polymer phase and a critical
point in-between pure gravity and polymers.  To
understand the behavior of strings in dimensions
higher than one, we need to obtain information
about the general critical properties of the effective
action in \basicint.  Since $R^2$ is a	hopping parameter,
we can write the hopping term as
$
L_{\mu}=\phi \Gamma_{\mu} \phi \Gamma_{\mu}^{\dagger}$
and in analogy with standard lattice analysis, we could
write an approximation to the effective action
in terms of a sum over connected graphs of angular
averages over the hopping terms generating the graph:
$$
\Gamma_{\rm eff} = \sum_{ C,{\rm Connected}}
\langle Tr \prod_{l\in C} L(l)\rangle _U,
$$
where the subscript $U$ indicates that the average is
taken only over the angular variables.	To every connected
graph the average over angles associates a set of products of
traces.  We can interpret these traces as different ways
of fracturing and breaking the loop $C$.  They
describe how the embedded surface is collapsed into
pieces.  If we add a local curvature term (the $A$-matrix),
we will obtain more general operators.	Since we do not have
an explicit way of evaluating \basicint\ and since we
are interested in the critical behavior of \raction\ and
not in the fine details of lattice dynamics, it is worth
exploring the type of universality classes of potentials
with arbitrary numbers of products of traces.  There are
two qualitative classes of potentials, depending on whether
they contain a finite or infinite number of traces.  So far
we have analyzed the case of
finite numbers of traces.

Defining
$$
t_i={1\over N} Tr\phi^{2i},
\eqn\trdef
$$
we take the effective action to be a general function
of an arbitrary, but finite number of $t_i$ variables,
$$
\Gamma_{\rm eff}=V(t_1,t_2,\ldots,t_n)=
V({1\over N} Tr\phi^{2},{1\over N} Tr\phi^{4},\ldots
{1\over N} Tr\phi^{2n}).
\eqn\potdef
$$
Thus we are faced with the analysis of the critical properties
in the planar limit of the action
$$
\eqalign{S(\phi)&=N^2 V(t_i)+N TrV_0(\phi),\cr
V_0&=\sum_k g_k Tr\phi^{2k}.\cr}
\eqn\barbonaction
$$
The world-sheet cosmological constant is related to
the the coupling $g_2=e^{-\mu_B}$.
We have made the simplifying assumption
of restricting our considerations to the case of even
potentials.  This is just a question
of technical simplicity, and
nothing changes if we take general potentials.
The analysis can be carried out for any of the Kazakov
potentials
\REF\kazcrit{V.A. Kazakov, Mod. Phys. Lett. {\bf A4} (1989) 2125.}
[\kazcrit].

The solution of the planar approximation of \barbonaction\ uses the
Hartree-Fock (HF) approximation, which becomes exact in the planar
limit. We first write down the saddle-point equations for
\barbonaction, and then we replace the traces \trdef\ by their planar
vacuum expectation values $x_i\equiv \langle {1\over N} Tr \phi^{2i}
\rangle$. This reduces the problem to the original
pure gravity case studied in [\brezin]. Finally we fix the variables
$x_i$ self-consistently.  The saddle-point equations and the HF
conditions are given by
$$
\eqalign{
\onet \sum_p 2p {\tilde g}_p \lambda^{2k-1} &={\rm P.V.}
\int d\mu{\rho(\mu)\over \lambda  - \mu},\cr
{\tilde g}_p &= g_p+{\partial V(x)\over \partial x_k},\cr}
\eqn\saddle
$$
$$
x_k=\int \rho(\mu) \mu^{2k}.
\eqn\hfcond
$$
$\rho(\mu)$ is the density of eigenvalues.
We will denote by ${\tilde V}$ the potential with the
shifted couplings
\saddle.

Following the pure gravity case, we look for
one-cut solutions of \saddle.  This means that the loop
operator
$$
F(p)=\int d\mu {\rho(\mu)\over p-\mu}
$$
is given by
$$
F(p)=\onet {\tilde V}'(p)-M(p)\sqrt{p^2-R}.\eqn\fpmp
$$
The polynomial $M(p)$ is determined by requiring that at
large values of $|p|$, $F(p)\sim 1/p + {\cal O} (1/p^3)$.  Since
$V$ depends only on a finite number of traces, the
potential ${\tilde V}$ is a polynomial, and there are
no ambiguities in the determination of $M$.

Define
$$
G(t)=pF(p)|_{p^{-2}=t} = 1+x_1(R)t + x_2(r) t^2+\cdots.
$$
The condition $x_0=1$ is the string equation.
Since $G(t)$ is analytic around $t=0$,
$$
\oint_{\cal C} t^l G(t) (1-tR)^{-1/2} =0, \eqn\intg
$$
for $l\ge0$ and ${\cal C}$ a circle around $t=0$ with radius
less then $1/R$.
Using \fpmp\ and
$$
{1\over \sqrt{1-Rt}} = \sum_{k\ge0} \bco k\left({Rt\over4}\right)^k,
\eqn\roote
$$
the equation \intg\ results in $M(p)=\sum_{m\ge0} M_m p^m$ with
$$
M_m = \sum_{k\ge0} \bco k\rfour k (m+k+1)\gtil {m+k+1}.\eqn\mms
$$
The density of eigenvalues is given by
$$
\rho (\lambda )={1\over 2\pi} M(\lambda)\sqrt{R-\lambda^2},
\qquad \lambda \in [-\sqrt{R},\sqrt{R}].\eqn\dene
$$

In the following we will also need expressions for $x_p(R) =
\langle {1\over N} Tr \phi^{2p}\rangle$. To derive those, we find it
convenient  to distinguish between two sources of $R$-dependence: the
explicit  $R$-dependence that would be here even for pure gravity,
\ie\
for $\gtil k=g_k$, and the implicit $R$-dependence of $x_p(R)$ (and
therefore of $\gtil k$), generated by the HF feedback. Consequently,
we will use two kinds of derivatives w.r.t. $R$: the usual derivative
$\partial/\partial R\equiv \partial_R$, taking into account both
sources of  $R$-dependence, and $D_R$,	derivative w.r.t. $R$ carried
out while ignoring the implicit  HF dependence of $x_k$. In
particular, $$
D_R G(t) = {1\over \sqrt{1-Rt}}\left[{M(1/t)\over2}-
D_RM(1/t)\left({1\over t}-R\right)\right].\eqn\drg
$$
Since $\sqrt{1-Rt} D_RG(t)$ is a series in positive powers of $t$, the
only term  from the square brackets on the r.h.s. of \drg\ that
contributes is $M_0/2+RD_RM_0$. Thus,
$$
\eqalign{D_Rx_p &= \oint t^{-p-1}{{M_0\over2}+RD_RM_0\over
\sqrt{1-Rt} },\cr
&=\left({M_0\over2}+R D_RM_0\right)\bco p\rfour p,\cr
&=\bco p\rfour p D_Rx_0.\cr}\eqn\drxp
$$
Finally, using the explicit expression for $M_0$ from \mms\ we arrive at
$$
x_p(R) = \sum_{k\ge 1}{k^2\over p+k} \bco p\bco k\rfour {k+p} \gtil k.
\eqn\xsubp
$$

The string equation is given by \xsubp\ with
$p=0$.	We are interested in the analysis of the
critical points
of \barbonaction, the computation of string susceptibilities, and
in the computation of correlation functions of
macroscopic loop operators.  If we denote by $\beta$ the coupling
$g_2$, the string susceptibility is determined from
$$
\chi ={dx_2\over d\beta}={\partial x_2\over \partial \beta}
+{dR\over d\beta} {\partial x_2\over \partial R}.
\eqn\stringsus
$$
This is because the coupling $\beta$ multiplies the operator
$t_2$ in the action. Therefore
$\partial \ln Z/\partial \beta$
is proportional to the expectation
value $x_2$, and the second
derivative of the logarithm of the partition function is the
string susceptibility.

The dependence of $R$ on $\beta$ is read off from the string
equation $x_0=1$.  In particular,
$$
0={dx_0\over d\beta}={\partial x_0\over \partial \beta}
+ {dR\over d\beta}{\partial x_0\over \partial R}
$$
allows us to read the equation determining the critical
points,
$$
{d\beta \over dR}=-{\partial_R x_0 \over
\partial_{\beta}x_0 },
\eqn\critpoint
$$
Since the denominator in \critpoint\ is proportional to
$R^2$, the critical points are determined by the zeroes
of the numerator.  We can similarly write
$\chi$ according to
$$
\chi =\partial_{\beta} x_2 - \partial_{\beta} x_0
{\partial_R x_2\over \partial_R x_0}.
\eqn\newsus
$$

The singularities in the behavior of $\chi$ come
from the ratio of derivatives in \newsus.  For the
standard Kazakov critical points [\kazcrit] \newsus\
does not blow up at the critical point $R_c$, and
$\chi-\chi_c\sim R-R_c$. Since at the $m$-th
critical point $\beta-\beta_c\sim (R-R_c)^m$
this yields $\gst =-1/m$.  As we will show presently,
the critical points we describe have $\gst >0$ and $\chi$
will be singular at the critical point.  All
the phase transitions are second order.

\def\ap{{2p\choose p}\left({R\over 4}\right)^p}
\def\ak{{2k\choose k}\left({R\over 4}\right)^k}
\def\al{{2l\choose l}\left({R\over 4}\right)^l}
\def\dr{D_R}

\def\parr{\partial_R}

To derive the criticality conditions, use \xsubp\ to write
the complete derivative of $x_p$ w.r.t. $R$ as
$$
\eqalign{\parr x_p=&\dr x_p +\sum_{k\ge 1} {k^2\over k+p}
{2k\choose k}{2p\choose p}\rfour {k+p} \parr \gtil k\cr
=& \dr x_p +\sum_{k\ge 1} {k^2\over k+p}
\ap \ak\sum_q\partial_{kq}V \parr x_q.\cr}
\eqn\totalr
$$
Defining the matrix
$$
U''_{pq}=\sum_{k\ge 1} {k^2\over k+p}
\ap \ak \partial_{kq}V,
\eqn\udef
$$
we can compute $\parr x_p$ in terms of
$\dr x_0$:
$$
\parr x_p=[(1-U'')^{-1} A]_p \dr x_0,
\eqn\dofxp
$$
where $A$ is a vector whose $p$-th component
equals $\ap$. When $p=0$, we obtain the criticality
condition implied by \critpoint:
$$
\parr x_0=[(1-U'')^{-1} A]_0 \dr x_0=0.
\eqn\critcon
$$

Notice that the criticality condition splits into
two terms. The first one is related to the
function including the couplings with multiple
traces,
$$
[(1-U'')^{-1} A]_0=0,
\eqn\typeone
$$
and the second is equivalent to the criticality
condition for the Kazakov critical points:
$$
\dr x_0=0.
\eqn\typetwo
$$
The term in \newsus\ which may lead to singularities
in $\chi$ and to positive $\gst$
is given by
$$
{\parr x_2\over \parr x_0}=
{[(1-U'')^{-1} A]_2\over [(1-U'')^{-1} A]_0}.
\eqn\suscrit
$$

In \critcon\ there are three possibilities: 1). The
polymer couplings in $U$ become critical but
the gravity part does not.  The zeroes of
$d\beta/dR$ come only from the polymer contribution.
 2). The gravity contribution is the only
one generating zeroes of $d\beta/dR$.  3). Both
terms become critical.	In the first case we
have a theory of polymers, and little is remembered
of the coupling to gravity. In the second case
the polymer degrees of freedom are completely
frozen and we reproduce the string susceptibility
and exponents of the Kazakov critical points.  The
third and more interesting case is when both the
``polymer" matter and gravity become critical
simultaneously.  This is the case where we can
find novel behavior.  Notice also that unless we
tune the numerator in \suscrit\ we will generically
obtain $\gst >0$ in cases 1) and 3).

If
\def\rc{R_c}\def\bec{\beta_c}
$\rc,\bec$ are the critical values of
$R$ and $\beta$, and if near the critical point
$$
[(1-U'')^{-1} A]_0\sim (R-\rc)^n\qquad
\dr x_0\sim (R-\rc)^m,
\eqn\nmcrit
$$
then
$$
\beta-\bec\sim (R-\rc)^{n+m+1}.
\eqn\bofrcrit
$$
If we do not tune the numerator of \suscrit\ to
partially cancel the zeroes in the denominator, we obtain
$$
\chi\sim{1\over (R-\rc)^n}.
$$
The previous two equation give a string susceptibility
for the $(n,m)$-critical point:
$$
\gst={n\over n+m+1}.
\eqn\nmstring
$$
By tuning the numerator in \suscrit\ we can change
the numerator in \nmstring\ to any positive integer
smaller than $n$
$$
\gst={p\over n+m+1}.
$$

We will show below that only in the case $n=1$ the model
maintains some resemblance with the properties one
would expect of a non-critical string.	We can
define operators creating macroscopic loop only
in that case.  For $n>1$ the polymerization of the
surface is so strong that there is no room left
to open macroscopic loops.

To illustrate the preceeding discussion, we present an
 explicit example. The simplest case to
study is the one where the function $V$ depends
on a single trace $V=V(t_l)$.  In this case the
matrix $U''$ becomes:
$$
U''_{pq}=\delta_{ql} U''_{pl}=\delta_{ql}
{l^2\over l+p}\ap\al V'',\qquad
p=1,2,\ldots.
$$
The matrix $U''$ contains a single non-vanishing
column in position $l$.  The inverse of $1-U''$ is:
$$
(1-U'')^{-1}= 1+{U''\over 1-{l\over 2}(\al)^2 V''},
$$
therefore,
$$
[(1-U'')^{-1}A]_p=
$$
$$
\ap +
{1\over 1-{l\over 2}(\al)^2 V''}{l^2\over p+l}
\left(\al\right)^2\ap V'',
$$
$$
[(1-U'')^{-1}A]_0=
{1+{l\over 2}(\al)^2V''\over
1-{l\over 2}(\al)^2V''}.
\eqn\xpxo
$$

The criticality condition \typeone\ is now
$$
1+{l\over2}{\bco l}^2 \rfour {2l} V'' \propto (R-R_c)^n.\eqn\lcrit
$$
We may simplify the expressions for the moments by eliminating
$V'$  from the string equation $x_0=1$ to get
$$
x_k={1\over l+k}\ak\left[l+\sum_p{pk(p-l)\over p+k}\ap g_p\right].
\eqn\xsubk
$$
The analogue of the criticality condition \typetwo\ is therefore
$$
RD_R x_0 = l+\sum_p p(p-l)\ap g_p \propto (R-R_c)^m.\eqn\rdrxo
$$
Note that we still need the string equation for $l \neq 2$, since
$\beta = g_2$ appears on the right hand side of these expressions.
Hence, the HF feedback is explicitly solvable only in the case $l=2$. On
the other hand, direct differentiation of \xsubk\ gives
$$
{\partial x_k \over \partial R} = {k\over l+k} A_k D_R x_0.\eqn\dxkdr
$$

This means that the total derivative $dx_k/dR$ is proportional to the
criticality condition \rdrxo, thus restricting the critical points
accessible
with the  simple potentials $V(t_l)$ to $(n,m=0)$ or $(n=1,m)$.
The criticality condition \lcrit\ for $n=2$ implies
$$
\eqalign{{dW\over dR}\biggl|_{R=R_c} &= l^2 \left(
\bco l\left({R_c\over 4} \right)^l \right)^2
V''(x_l(R_c))+ {l\over 2}\left(\bco l\left({R_c\over 4} \right)^l
\right)^2 V'''(x_l(R_c)) {dx_l\over dR}\biggl|_{R=R_c},\cr
&= -2l+{l\over 4}\left(\bco l\left({R_c\over 4} \right)^l \right)^3
V'''(x_l(R_c)) D_R x_0(R_c) = 0,\cr}
$$
where $W\equiv {l\over 2}\left(\bco l\left({R\over 4} \right)^l
\right)^2 V''(x_l(R))$. Obviously, the last equation is inconsistent
with $D_Rx_0(R_c) =0$, the condition for $m>0$.
One
can show, following the same line of reasoning that we used here, that a
more general class of potentials with $V=V(\sum \alpha_r
t_r)$ can accomodate $n>1$, $m>0$ points, with a judicious choice of
$\alpha_r$'s. Since
the
physically  interesting case will anyhow turn out to be $n=1$, we will
relegate
the discussion of this more general class of potentials to appendix A.

Finally, let us display a few explicit potentials resulting in a
critical behaviour of the new kind that we are studying. Specializing
to $V=V(t_2)$, the criticality conditions \typeone\ and \typetwo\
become
$$
\eqalign{1+36\rfour 4 V''(x_2) &\propto x^n,\cr
2+\sum_{p\ge 1} p(p-2) g_p\ap &\propto x^m,\cr}\eqn\ltwocrit
$$
where $R=R_c(1+x)$ and $x\rightarrow 0$ is the critical point. Another
useful identity is
$$
x_2 = 3\rfour 2+3\sum_{p\ge1} {p(p-2)\over p+2} g_p \bco p\rfour {p+2}.
\eqn\xtwo
$$

It will be sufficient to look at the potential
$$
U=g_1 t_1+\beta t_2 + h t_2^2.
$$
We choose $R_c=4$. To obtain $n=1$, $m=0$ critical point all we need
 now is $h=-1/72$. The string equation $x_0=1$ fixes $\beta=1/18+x^2/6+
{\cal O}(x^3)$ which implies $\gamma=1/2$. The eigenvalue density
\dene\ is (neglecting ${\cal O}(x^2)$ in $M(\lambda)$),
$$
\rho(\lambda)= {1\over\pi}\left( {1\over2}- {x\over3}-{x\over 6}
\lambda^2\right) \sqrt{4(1+x)-\lambda^2},
$$
which for $x\rightarrow 0$ reduces precisely to Wigner's semicircle
law . For $n=m=1$ \ltwocrit\ gives $h=-1/72$ and $g_1=1$, whereas
$x_0=1$
gives $\beta=-1/18-2x^3/9 + {\cal O}(x^4)$ implying $\gamma=1/3$.
The critical eigenvalue density  is exactly equal to the pure gravity
one ($\gamma= -1/2$)
$$
\rho(\lambda)={1\over 6\pi}(4-\lambda^2)^{3/2}.
$$
Note that the critical eigenvalue distribution reflects only the
Kazakov part of the criticality conditions, while it is insensitive to
\typeone, the condition responsible for divergence of $\chi$ and for
positive string  susceptibility.

An argument similar to the one presented at the end of
[\tata] shows that the critical point for the transition
between the polymer and the pure gravity phases
is generically second order. To see this, we look at the
partition function for "daisies" at a fixed number of Feynmann
vertices (or elementary plaquettes in the dual version). We
can extract this quantity from the asymptotics
$$
Z \sim \sum_n  n^{\gst -3}\left({\beta\over \beta_c}\right)^{n}
$$
So that, in the thermodinamic limit $n\rightarrow \infty$ we have
the free energy
$$
f = {\rm log}|\beta_c|
$$
where $\beta_c(g, V)$ is the convergence radius of the planar
expansion as a function of the couplings in the matrix-model
Lagrangian. $g$ denotes collectively the linear couplings except
$g_2 = \beta$, while $V$ corresponds to non-linear terms like
$(Tr \phi^{2})^{2}$. We are interested in singularities of
$\beta_c$ as a function of $V$, since these couplings act as the
"temperature" for the gas of splitting points. A first order
phase transition would imply coexistence of the smooth and polymer
phases, while a second order transition represents global condensation
of daisies at the critical point. We calculate the function
$\beta_c (g, V)$ as
$$
\bec (g, V) = \beta ( \rc(g, V)), g, V)\eqn\vayavaya
$$
where $\beta (R, g, V)$ is the solution to the s self-consistency
conditions (equations for the moments $x_k$) plus the string equation.
It depends on the planar solution and it is a regular function of
the couplings $g, V$. $\rc (g, V)$ is the position of the critical
point as a function of the couplings, and we can find it from the
criticality conditions \critcon:
$$
{\partial \beta\over \partial R}(\rc, g, V) = 0 \eqn\una
$$
Tunning $g$ and $V$ to some submanifold of appropriate codimension
we can arrange zeroes or any order in \una. Further change of $g, V$
within
these universal submanifolds amounts simply to move the location of
the critical point $\rc$. In our case \una\ has a factorized structure
as in \critcon, and we have two branches of
critical points: $\rc^{I}$ and $\rc^{II}$, determining two functions
in \vayavaya; $\bec^{I}$ and $\bec^{II}$. The smaller of these values
determines the radius of the covergence disk for the planar
perturbation series around $\beta=0$. Hence we are interested in the
behaviour
of the function
$$
\bec (g, V) ={\rm min}\{|\bec^{I}|, |\bec^{II}|\}
$$
near the coalescing
region $\rc^{I} \sim \rc^{II} \sim \rc^{*}$, reached by tunning $V$.

$\bec (g, V)$ is continous across the transition, and the first
derivative is
$$
{d\bec\over dV}(g,V) = {\partial \beta\over \partial V}(\rc,g,V)
 + {\partial \beta\over \partial R}(\rc(g,V), g,V){\partial \rc
\over \partial V}(g,V)
$$
The first term on the right hand side is continous.
In the second term, $\partial
\rc \over \partial V$ may have some insolated singularities, but we
can assume smooth behaviour in open sets. On the other hand
${\partial \beta\over \partial R}(\rc)$ vanishes identically for
all critical points, not only for the coalescing one. So the
second term is zero along the critical submanifold and $d\bec\over
 dV$ is continous at $V^{*}$.

The second derivative contains unprotected terms like $ {\partial
\beta\over \partial R \partial V}(\rc(g,V),g,V)$. These are
generically discontinous since $\rc^{I}$ depends on the first
derivative of the potential, while $\rc^{II}$ depends on the
second derivative. Thus we are led to at least second order phase
transitions and a continuum limit is possible as expected.

We have presented the general analysis of effective actions
in the planar limit containing arbitrary, but finite
numbers of traces \barbonaction.  We were able to show
that beyond some critical coupling, $\gst$ becomes positive
and takes values of the form $\gst=n/(n+m+1)$ for any positive
integers $n,m$.  We should continue exploring the
properties of these critical points and their scaling
operators.  Obviously the more challenging case of studying
arbitrary functions $V$ depending on an infinite number
of traces is crucial before we can draw any conclusions
concerning the properties of our model.  This together
with the study of general properties of \basicint\
should shed further light into the properties of non-critical
strings in interesting dimensions $D=2,3,4$.  It is
very important to obtain in what way the effective actions
studied in the previous sections depend on dimensionality.

\chapter{MACROSCOPIC LOOPS}

To explore the new critical points in some detail, we
compute the correlation functions of macroscopic
loop operators.  The simplest loops to compute are
the ones analogous to those appearing in pure gravity
(see for instance
\REF\mss{G. Moore, N. Seiberg and M. Staudacher, \npb
{\bf B 362} (1991) 665.}[\mss]).  They are obtained by taking
the limit $k\rightarrow\infty$ of $Tr\phi^{2k}$.  In
terms of the original reduced model these are loop on
the surface whose position is averaged over the target
space.	One could also define loops with definite
positions in the target space.	The expectation values
of the latter would certainly give crucial information
on the reduced formulation of string theory, but we
have not yet calculated them.  For the simpler, pure
gravity loops of length $l$ we take the definition
$$
\langle w(l)\rangle=\lim_{k\rightarrow\infty}
\sqrt{{\pi\over l}}\langle Tr \phi^{2k}\rangle =
N\lim_{k\rightarrow\infty}\sqrt{{\pi\over l}} x_k.
\eqn\lloop
$$
As a matter of convenience we choose the critical values
for $\beta$ and $R$ as $\bec=-1, R_c=1$.  In the
Kazakov critical points, $\gst=-1/m$ and the scaling
limit is taken to be $Na^{2-\gst}=\kappa^{-1}$.  It is
a simple but non-trivial fact that the correct way to
define a loop of length $l$ in this case is to scale
$k$ according to $ka^{2\gst}=l$ as $k\rightarrow\infty$
(see for example [\mss] and references therein).  In our
case we will restrict for simplicity
our computations to the case of
potentials $V=V(t_l)$.

We will start by calculating $dx_k/dg_j$. From \xsubp\ we have
$$
\eqalign{{\partial x_l\over\partial g_j} &= {{j^2\over j+l}
\bco l\bco j\rfour{j+l}\over 1-W},\cr
{\partial x_k\over\partial g_j} &= {j^2\over j+k}
\bco k\bco j\rfour{j+k}{1-{(k-l)(j-l)\over(k+l)(j+l)}W\over 1-W},\cr}
\eqn\dxdgj
$$
where we have introduced the notation $W\equiv {l\over2}\left(\bco l
\rfour l\right)^2 V''(x_l)$. From \xpxo\ we have
$$
{\partial_R x_k\over \partial_R x_o} = \bco k\rfour k
{1-{k-l\over k+l} W \over 1+W}.\eqn\dxow
$$
Putting it all together,
$$
\eqalign{{dx_k\over dg_j} &= {\partial x_k\over \partial g_j}-
{\partial x_0\over \partial g_j}
{\partial_R x_k\over \partial_R x_0},\cr
&= -{jk\over j+k}\bco k\bco j\rfour{k+j}+{jk\over (j+l)(k+l)}
\bco k\bco j\rfour{k+j} {2lW\over 1+W}.\cr}\eqn\dxk
$$ For $V''=0$, up to irrelevant numerical factors this answer
coincides with the one-loop expectation value in the Kazakov critical
points.

To go to the continuum limit we have to choose the scaling
variables.  Recall that in the double scaling limit [\doubles] the
free energy takes the form $$
F=\sum_{g\ge 0}N^{2-2g} (\mu-\mu_c)^{(2-\gst)(1-g)} F_g
$$
where $\mu=-\log\beta$, and $g$ is the genus of the
triangulations summed over.  The renormalized cosmological
constant is introduced according to
$\mu-\mu_c=a^2 t$ where $a$ is the microscopic lattice
spacing.  Then the effective string coupling constant
is defined by $N a^{2-\gst}=\kappa^{-1}$.  Hence the
scaling variable associated with the string susceptibility
is $\chi = a^{-2\gst} u$.  At the $(n,m)$-critical point
$\chi$ behaves as $\chi\sim(1-R)^{-n}$.  Thus the
scaling variables are
$$
\chi\sim{1\over (1-R)^n}=a^{-2\gst}u,\qquad
1-R=a^{2\gst/n} u^{-1/n},
$$
$$
\beta-\beta_c=\beta+1=a^2 t.
\eqn\scalevar
$$
For an $(n,m)$-critical point we have
$$
1-R=a^{{2\over n+m+1}}u^{{-1\over n}},\qquad
\chi = u a^{-{2n\over n+m+1}},\qquad
N a^{2-{n\over n+m+1}}=\kappa^{-1}.
\eqn\critscaling
$$

An insertion of a loop operator is a string interaction which brings
in an extra power of $\kappa$, the string coupling constant. Therefore,
 in general we expect
$$
\langle w(\ell_1)\ldots w(\ell_M)\rangle_c =
\kappa^{M-2} \langle \hat w ( \ell_1)\ldots \hat w ( \ell_M)
\rangle_c,\eqn\trlm
$$
and in particular
$$
\langle w(\ell)\rangle={1\over\kappa}\langle\hat w ( \ell)\rangle.
\eqn\trlone
$$
This will allow us to arrive at the proper definition of the length
stick, \ie\ to fix $y$ in $\ell=ka^y$. Setting $j=2$ in \dxk\
($g_2\equiv\beta$, $1+W \propto (R_c-R)^n$), we easily obtain, up to
a non-universal positive multiplicative constant
$$
{d\over dt}\langle Tr \phi^{2k}\rangle = Na^2\left( {a^{y/2}\over
\sqrt{\pi \ell}}\left(1-a^{{2\over n+m+1}}
u^{-1/n}\right)^{\ell/a^y}
ua^{-{2n\over n+m+1}}\right).\eqn\contw
$$
Identifying the r.h.s. with ${1\over\kappa} {d\over dt}\langle\hat
 w(\ell)\rangle$ leads to
$$
\eqalign{\ell&\equiv ka^{{2n\over n+m+1}},\cr
{d\over dt} \langle w(\ell)\rangle &= -{u\over \kappa\ell}
e^{-\ell/u}\qquad {\rm for} \; n=1,\cr
\langle w(\ell)\rangle &=0 \qquad {\rm for}\; n>1.\cr}\eqn\onel
$$
In the phases with $n>1$ the
polymer couplings dominate completely the critical behavior,
in spite of the fact that gravity becomes also critical,
in such a way that there is not enough room to open
macroscopic loops.\foot{Strictly speaking, we should have
put this discussion in the framework of potentials with
 $V(\sum \alpha_r t_r)$, the simplest ones with $n>1$
critical points. As shown in the appendix A, the much more
complicated formulas in that case lead again to \contw\ and to
the same conclusion that $\langle w(\ell)\rangle=0$ for $n>1$. }

In the case of $n=1$ we can also calculate two-loop and multiloop
correlation functions. Let $\ell_1=ka^{{2n\over n+m+1}}$,
$\ell_2=ja^{{2n\over n+m+1}}$. For $n=1$, $1+W=2la^{{2\over m+2}}
u^{-1}$. For $k,j\rightarrow \infty$ and $a\rightarrow 0$, \dxk\
 gives the continuum two-loop operator
$$
\eqalign{\langle w(\ell_1)w(\ell_2)\rangle &\equiv
\lim_{k,j\rightarrow\infty}
{\pi\over\sqrt{\ell_1\ell_2}} \left\langle Tr\phi^{2k}
Tr\phi^{2j}\right\rangle_c,\cr
&={e^{-(\ell_1+\ell_2)/u}\over \ell_1+\ell_2}+
{u\over \ell_1\ell_2} e^{-(\ell_1+\ell_2)/u}.\cr}
\eqn\twoloop
$$
The lesson
we can draw from \twoloop\ is that in the simple case of
$n=1$ the effect of the polymer couplings is to contribute
an extra state (the last term in \twoloop\ ) which resembles
very much the contribution one would expect of a tachyon.
This term represents the breaking of the cylinder
interpolating between the two loop of lengths $l_1,l_2$
into two disks osculating at one point.  We can represent
the last term in \twoloop\ as
$$
\langle w(l_1)P\rangle {1\over \langle PP\rangle}
\langle P w(l_2)\rangle
$$
where $P$ is the puncture operator.

With \onel\ and \twoloop\ it is easy to obtain the multi-loop
correlators as well. We can write $\langle w(\ell_1)w(\ell_2)
\rangle$ in two ways:
$$
\langle w(\ell_1)w(\ell_2)\rangle=\lim_{k_1\rightarrow\infty}
\left(-{1\over N}\right){1\over\kappa \ell_2} \int\limits_t^\infty
\left(1+{\ell_2\over u}\right) {du\over dg_{k_1}} e^{-\ell_2/u} dt',
\eqn\wto
$$
and
$$
\langle w(\ell_1)w(\ell_2)\rangle=-{1\over \ell_1\ell_2}
\int\limits_t^\infty
\left(1+{\ell_1\over u}\right) \left(1+{\ell_2\over u}\right)
{du\over dt'} e^{-(\ell_1+\ell_2)/u} dt'.\eqn\wtt
$$
Comparing \wto\ and \wtt\ we have
$$
\lim_{k\rightarrow\infty} \sqrt{{\pi\over \ell}} \left(- {1\over N}
\right){du\over dg_{k}} =
-{\kappa\over\ell} \left(1+{\ell\over u}\right) {du\over dt}
e^{-\ell/u}.
$$
Define the operator of insertion of a macroscopic loop $D(\ell)$ as
$$
D(\ell)\equiv -{\kappa\over\ell}\sum_{p\ge 0}{d^{p}\over dt^{p}}
\left\{ \left(1+{\ell\over u}\right) e^{-\ell/u} {du\over dt}\right\}
{\partial\over\partial u^{(p)}}.
$$
Where $u^{(p)} = {d^{p}u\over dt^{p}}$. As it should be, $D(\ell)$ is
proportional to $\kappa$, and we can write the $M$-loop amplitude as
$$
\langle w(\ell_1)\ldots w(\ell_M)\rangle = \left(\prod_{i=2}^M
D(\ell_i)\right) \langle w(\ell_1)\rangle.
$$
The pattern is
reminiscent of the residue of the tachyon pole in the
Belavin-Knizhnik theorem
\REF\bk{A.A. Belavin and V. Knizhnik, \plb {\bf 168 B}(1986)
201; J.B. Bost and T. Jolicoeur, \plb {\bf 174 B}(1986) 273;
R. Catenacci, M. Cornalba, M. Martellini and C. Reina,\plb
{\bf 172 B}(1986) 328.}
[\bk].	This interpretation, though tempting, has to be
taken with several grains of salt because in the types
of effective actions we are studying it is hard
to see any dependence on dimensionality.  Nevertheless
we take this result as
encouraging evidence that our approximation captures
some of the expected properties of non-critical strings
for $D>1$.

The form of the two-loop operator \twoloop\ does not
depend on the form of the potential $V(t_1,\ldots, t_n)$
as long as we consider $(1,m)$-critical points.  Hence
only the quadratic part of $V$ determines the critical
properties of macroscopic loops.

To conclude this section we mention that we can
study, at least in part, the spectrum of the scaling operators
by perturbing the criticality conditions.  The string
equation is then modified according to
$$
1+\beta=(1-R)^{n+m+1}+\sum \tau_k^B (1-R)^{k+1}
$$
The subleading contributions depend on both the polymer
and Kazakov perturbations of the critical conditions
\critcon. The bare couplings of the scaling operators
are the $\tau_k^B$ parameters.	Using \critscaling\ the string
equation becomes
$$
t=u^{-{n+m+1\over n}}+\sum_k \tau_k u^{-{k+1\over n}},
\eqn\stringeq
$$
and the $\tau_k$ are the renormalized couplings.  In the
original model these couplings would hardly exhaust the
scaling operators, although they may represent a
significant subset associated to the
simplest loop operators.  If we call $\sigma_k$ the
scaling operator associated to the coupling $\tau_k$, its planar
correlators at the $(n,m)$-critical point follow from
\stringeq\
$$
{du\over d\tau_k}=-u^{-{k+1\over n}}{du\over dt}
$$
and
$$
\langle \sigma_k PP\rangle={n\over n+m+1}
t^{{k-m-2n\over n+m+1}}.
$$

\chapter{$1/N^2$-CORRECTIONS, DISCRETE AND CONTINUUM LOOP EQUATIONS}

In this section we will use Schwinger-Dyson equations to evaluate
$1/N^2$-corrections to our planar results. The $1/N^2$-corrections
will prove to be consistent with string perturbation expansion and
with double scaling. \foot{Since the equivalence of the original
action \masterone\ and its reduced version \rstring\ has been
demonstrated only in the planar approximation, the $1/N^2$-corrections
for \barbonaction\ are one step further removed from \masterone.
Still, in order to be able to take the new critical points of our
effective action \barbonaction\ more seriously, we want to see if
their continuum limit is consistent with a string theory.}
In particular, we will calculate the string
susceptibility on the torus to be
$$
\gamma_{{\rm torus}} = +2.
$$
It is amusing to note that this is in agreement with DDK formula
$\gamma_h = 2+(1-h)\gamma_0$ for string susceptibility at genus $h$,
where $\gamma_0$ (a complex number for $c>1$) drops out at $h=1$.

We will study matrix integrals of the form
$$
\eqalign{Z &=\int {\cal D}\phi \exp\left\{-N^2\left[\sum_{n=1}^s
g_n\left({1\over N} Tr \phi^{2n}\right)+V(t_2)\right]\right\},\cr
&\qquad V(t_2) = \sum_{m=2}^t h_m\left({1\over N} Tr
\phi^4\right)^m.\cr}
$$
The structure of the calculation follows \REF\ambmak{J. Ambj\o rn and
Yu. M. Makeenko, Mod. Phys. Lett. {\bf A5}, (1990) 1753.} [\ambmak].
In a standard fashion,
$$
0=\int {\cal D}\phi {d\over d\phi_{ab}} \left\{(e^{\ell\phi})_{ab}
\exp\left\{-N^2\left[\sum_{n=1}^s
g_n\left({1\over N} Tr \phi^{2n}\right)+V(t_2)\right]\right\} \right\}
$$
leads to the loop equation
\def\overn{{1\over N}}
\def\leftl{\left\langle}
\def\rightr{\right\rangle}
\def\tre#1{Tr e^{#1\phi}}
\def\intol{\int\limits_0^\ell}
$$
\eqalign{
\left\langle \int\limits_0^\ell d\ell' \overn\tre{\ell'}\overn
\tre{(\ell-\ell')}\rightr &= \sum_{n=1}^s g_n 2n \leftl\overn
Tr\left(\phi^{2n-1} e^{\ell\phi} \right)\rightr \cr
&\quad + \sum_{m=2}^t h_m 4m\leftl \left(\overn Tr\phi^4\right)^{m-1}
\overn Tr\left(\phi^3 e^{\ell\phi}\right) \rightr.\cr}
\eqn\loopeq
$$

Let us define connected correlation functions $\Phi_{\ell_1\cdots
\ell_M}$ as
$$
\Phi_{\ell_1\cdots \ell_M}\equiv N^{M-2} \leftl \tre{\ell_1}\cdots
\tre{\ell_M} \rightr_c,
$$
\def\sumones{\sum_{n=1}^s}
where the explicit factor of $N$ makes all $\Phi_{\ell_1\cdots
\ell_M}$ of order one. For $N$ large, \loopeq\ leads to
$$
\eqalign{\sumones 2n\gtil n \left({d\over d\ell}\right)^{2n-1} &\leftl
\overn \tre\ell\rightr =\intol d\ell'\leftl\overn \tre{\ell'} \rightr
\leftl\overn \tre{(\ell-\ell')} \rightr \cr
 +{1\over N^2}\biggl[\intol d\ell' &\leftl \tre{\ell'}
\tre{(\ell-\ell')} \rightr_c - 4V''(x_2) \leftl Tr\phi^4 Tr\left(
\phi^3 e^{\ell\phi} \right)\rightr_c \cr
& -2 V'''(x_2) \Phi_{4,4} \leftl \overn Tr\left(\phi^3
e^{\ell\phi}\right) \rightr \biggr] + {\cal O}(1/N^4),\cr}
\eqn\aploop
$$
where
$$
\eqalign{\gtil n&=g_n + V'(x_2) \delta_{n,2},\cr
x_i &= \leftl\overn Tr\phi^{2i}\rightr,\cr
\Phi_{n,m} &= \leftl Tr \phi^n Tr\phi^m\rightr_c.\cr}
$$

After the Laplace transform, \aploop\ becomes
$$
\eqalign{
\tilde V'(p) G(p) = Q(p) &+ G^2(p)+ {1\over N^2}\biggl[ \Phi(p,p) -
4V''(x_2) (p^3 \Phi_4(p) - \Phi_{2,4}) \cr
&-2V'''(x_2)\Phi_{4,4} (p^3 G(p) -p^2 -x_1)\biggr] +{\cal
O}(1/N^4).\cr}\eqn\laploop
$$
We have introduced the notations
$$
\eqalign{\tilde V'(p) &= \sumones 2n\gtil n p^{2n-1},\cr
Q(p) &= \sumones 2n\gtil n\sum_{i=0}^{n-1} p^{2n-2i-2} x_i,\cr
G(p) &= \int\limits_0^\infty d\ell e^{-\ell p}\leftl \overn \tre\ell
\rightr = \leftl\overn Tr{1\over  p-\phi}\rightr,\cr
\Phi(p,p') &= \int\limits_0^\infty d\ell\int\limits_0^\infty d\ell'
e^{-\ell p}e^{-\ell' p'} \Phi_{\ell,\ell'},\cr
\Phi_4(p) &= \leftl Tr\phi^4 Tr{1\over	p-\phi}\rightr_c =
\sum_{n=1}^\infty \Phi_{4,2n} p^{-2n-1}.\cr}
$$

The solution to \laploop\ is well-known in the planar limit:
$$
\eqalign{G^{(0)} (p)
&= {1\over 2} \tilde V'(p) - M(p)\sqrt{p^2-R},\cr
&= \sumones n\gtil n p^{2n-1} - \sumones p^{2n-2}
\sum_{k=0}^{s-n} \ak (k+n)\gtil{k+n} \sqrt{p^2-R},\cr}
$$
where $R$ is fixed by the string equation
$$
x_0=1=\sumones n\gtil n\bco n\rfour n.
$$
To include $1/N^2$-corrections, write
$$
\eqalign{\tilde V' &=\tilde V'^{(0)} + \delta \tilde V' =
\tilde V'^{(0)} +4V''(x_2) \delta x_2 p^3,\cr
G &= G^{(0)} + \delta G,\cr
Q &= Q^{(0)} + \delta Q,\cr}
$$
where $\delta(\;)$ is ${\cal O}(1/N^2)$. From \laploop\ we obtain
the equation for $1/N^2$-corrections
$$
\eqalign{4&V'' (x_2) \delta x_2 p^3 G^{(0)}(p) + 2M^{(0)}(p)
\sqrt{p^2-R} \delta G(p)
 = \delta Q(p) +\cr
&+ {1\over N^2}\biggl[ \Phi(p,p) - 4V''(x_2)(p^3
\Phi_4(p) -\Phi_{2,4})
 - 2V'''(x_2) \Phi_{4,4}(p^3 G(p)
-p^2-x_1)\biggr]^{(0)},\cr}\eqn\nseq
$$
where
$$
\delta Q(p) = \sum_{n=2}^s 2n\gtil n\sum_{i=1}^{n-1} p^{2n-2i-2}
\delta x_i + 4V''(x_2)\delta x_2(p^2+x_1).
$$

An important piece of information that we can extract from \nseq\ is
the dependence of $\delta x_2$ on $R-R_c$. This will give
$1/N^2$-corrections to specific heat. To simplify \nseq\ we use the
fact that, for $s\ge 2$, $M^{(0)}(p)$ vanishes for $p$ equal to some
$p_0$. Therefore, at $p=p_0$ \nseq\ reduces to
$$
\eqalign{4V''(x_2) \delta x_2 p_0^3 G^{(0)}(p_0) = \delta Q(p_0)
+{1\over N^2}\biggl[&\Phi(p_0,p_0) - 4V''(x_2)(p_0^3
\Phi_4(p_0) -\Phi_{2,4})\cr
& - 2V'''(x_2) \Phi_{4,4}(p_0^3 G(p_0)
-p_0^2-x_1)\biggr]^{(0)}.\cr}
\eqn\poloop
$$
We need the planar expressions for $\Phi(p,p)$ and $\Phi_4(p)$. They
are calculated in appendix B. The results are
$$
\eqalign{\Phi_4(p) &={1\over 1+W}\left[{8p^4-4Rp^2-R^2\over
4\sqrt{p^2-R}} - 2p^3\right],\cr
\Phi(p,p) &= {R\over 4(p^2-R)^2}\cr
&\; -{4W\over 9R^4(1+W)}\left[
128p^6+ {16p^3 (R^2+4Rp^2-8p^4)\over \sqrt{p^2-R}} + {R^3(8p^2+R)\over
p^2-R}\right].\cr}
\eqn\chis
$$

We are ready to look at some explicit examples. As discussed earlier,
$V(x_2)$ potentials can describe critical points with either $m=0$ or
$n=1$. The simplest non-trivial case are potentials with $s=2$ which
allow us to look at $m=0$, $n\ge1$ and $m=n=1$ points.

For $s=2$ \poloop\ gives
$$
\eqalign{&(p_0^3 G(p_0) - p^2_0 - x_1)\left(4V''(x_2) \delta x_2 + {2\over
N^2} V'''(x_2) \Phi_{4,4}\right) =\cr
&\; = 4\gtil 2 \delta x_1+{1\over N^2}\left[\Phi(p_0,p_0) - 4V''(x_2)
(p_0^3 \Phi_4(p_0) - \Phi_{2,4})\right].\cr}
$$
Another relation between the unknowns $\delta x_1$ and $\delta x_2$
comes from the coefficient of $1/p^2$ in \nseq:
$$
\delta x_1 = -{2\over \gtil 1}(\gtil 2+V''(x_2) x_2)\delta x_2 -
{1\over N^2} {\Phi_{4,4}\over \gtil 1}(2V''(x_2) + V'''(x_2) x_2).
$$
Finally,
$$
\delta x_2 = {1\over N^2} {{\cal N}\over {\cal D}},
$$
where
$$
\eqalign{{\cal N}&\equiv
{R\over 4(p^2_0-R)^2} - {9R^4\over
16(1+W)}\left[- {3R\gtil 2 V''(x_2)\over \gtil 1(p^2_0-R)} +
V'''(x_2)\left({\tilde G(p_0)\over 2}+{\gtil 2\over \gtil 1}
x_2\right)\right],\cr
{\cal D} &\equiv
4V''(x_2) \tilde G(p_0) + {8\gtil 2\over \gtil
1}(\gtil 2+V''(x_2) x_2),\cr
\tilde G(p_0) &\equiv p_0^3 G(p_0) - p^2_0 - x_1 = \gtil 1 p^4_0 +
2 \gtil 2 p^6_0-p^2_0-x_1,\cr
p^2_0 &= -{R\over 2} - {\gtil 1\over 2\gtil 2}.\cr}
$$

Specializing further to $m=0$, $n\ge 1$ potentials with
$\gamma_0=n/(n+1)$, we set $R=4(1+x)$, $x\ll1$ and $\gtil 1=1/2$. From
$x_0=1$ we get $\gtil 2=-x/(12(1+x)^2)$ and $p^2_0 =3/x+{\cal O}(1)$.
Note that $\gtil 2$ is ${\cal O}(x)$ and $p^2_0$ diverges as we
approach the critical point, which is as it should be since $m=0$
potentials have Gaussian linear part of the potential (\ie\ $\gtil
n=0$, $n>1$).

Setting finally $h_2=-1/72$, $h_m=0$, $m>2$ we have fully specified
the
$m=0$, $n=1$ potential. We have ${\cal N} = -2x/3+{\cal O}(x^2)$ and
${\cal D}=-4x^3/9+{\cal O}(x^4)$. Furthermore, $\beta =
1/18+x^2/6+{\cal O}(x^3)$. Therefore,
$$
\delta x_2\propto {1\over N^2} {1\over x^2},\qquad {d\delta x_2\over
d\beta}\propto {1\over N^2}{1\over x^4} \propto {1\over N^2}{1\over
(\beta-\beta_c)^2}
$$
and
$$
\gamma_1 = +2.
$$

We have derived the first two terms in the genus expansion of the
specific heat. In terms of renormalized continuum variables we have
$$
{d^2\over dt^2} \log Z = \kappa^{-2}\left[{1\over t^{\gamma_0}} +
\kappa^2 {1\over t^2} +\cdots\right].
\eqn\spart
$$
The $1/N^2$-correction is consistent with string perturbation
expansion and with the double scaling limit.

One can repeat this calculation for $m=0$, $n>1$ and one is always led
to \spart. Similarly, for $n=m=1$ potential that we already discussed
we have $\gtil 1=1$, $h_2=-1/72$, $h_m=0$, $m>2$, $\gtil 2=-{1+2x\over
12(1+x)^2}$, $\beta=-1/18-2x^3/9 + {\cal O}(x^4)$, ${d\delta x_2\over
d\beta}\propto {1\over N^2}{1\over x^6}$ and we again have $\gamma_1
=2$ and \spart.

To reach $n=1$, $m=2$ we need potentials with $s=3$. Here $M(p)$ will
have two zeroes, $(p_0^\pm)^2$, that will both coalesce with $R$ at
$x=0$. If we take $\gtil 1=3/2$, $\gtil 3=1/60$, $h_2=-1/72$, $h_m=0$,
$m>2$ we obtain $\beta=-7/30 + x^4/4 + {\cal O}(x^5)$ and
$$
\eqalign{(p^+_0)^2 &= 4+(-1+\sqrt{-5})x -2\sqrt{-5} x^2 + {\cal
O}(x^3),\cr
(p^-_0)^2 &= 4+(-1-\sqrt{-5})x +2\sqrt{-5} x^2 + {\cal
O}(x^3).\cr}
$$
Now \nseq\ will result in two equations with two unknowns, $\delta
x_1$ and $\delta x_2$. Their solution again leads to $\gamma_1=2$ and
\spart.

It is interesting that there is a string perturbation expansion
\spart\ even for $n>1$. At those critical points, as we have seen
through vanishing of the macroscopic loop operators, polymerization
is so strong that it is impossible to open big holes in the surface.
Since the only continuum objects used in the derivation of \spart\ are
microscopic ones like $d/dt \sim P$, we were able to avoid pathologies
due to vanishing continuum macroscopic loop operators.

Finally, let us look at the continuum limit of the loop equation
\laploop\ itself. Comparing the solution of the continuum limit of the
loop equations with the continuum limit of the solution of the
discrete loop equations is a useful check of our calculations.

To begin with, use $\ell=ka^{{2n\over n+m+1}}$, $p=p_c + a^{{2n\over
n+m+1}} z$, to determine the scaling for the universal, continuum loop
correlation functions. On the one hand, we have
$$
\leftl \tre{\ell_1}\ldots \tre{\ell_M}\rightr_c = \kappa^{M-2}\int
dz_1\ldots dz_M\langle\hat w(z_1)\ldots \hat w(z_M)\rangle_c.
$$
On the other hand,
$$
\eqalign{
\leftl \tre{\ell_1}\ldots \tre{\ell_M}\rightr_c &=  {1\over N^{M-2}}
\Phi_{\ell_1\cdots\ell_M} =\cr
&= {1\over N^{M-2}} \int dp_1 \ldots dp_M \langle w(p_1)\ldots
w(p_M)\rangle_c.\cr}
$$
Consistency demands
$$
\langle w(p_1)\ldots w(p_M)\rangle_c = a^{{2(2m+n+2)-M(3n+2m+2)\over
m+n+1}} \langle\hat w(z_1)\ldots \hat w(z_M)\rangle_c,
$$
with $n\ge1$, $m\ge0$ and $M\ge0$. In particular,
$$
G(p)|_{\rm universal\> part} = a^{{2m-n+2\over m+n+1}} \langle \hat
w(z)\rangle.\eqn\gscale
$$
The negative sign of $n$ in the exponent is the first sign of problems
for $n>1$.

We start again with $n=m=1$ point. Set
$$
\eqalign{\beta &=\beta_c + a^2 t,\cr
R &= R_c(1+a^{2/3} t^{1/3}),\cr
\gtil 2 &= \beta+ 2h_2 x_2,\cr
p &= p_c + a^{2/3} z,\cr
\gtil 1 &= 1+c a^2 t.\cr}
$$
Demanding that universal (non-analytic in $t$) pieces in $x_1$ and
$x_2$ scale as $\langle P\rangle$ fixes $R_c=4$, $\beta_c+2 h_2 =
-1/12$. The criticality condition for $n=1$ gives $h_2=-1/72$,
$\beta_c =-1/18$. Finally, since we want to blow up the critical
region between the branch points, we fix $p_c^2=R_c=4$.

Plugging all of this into solution of discrete planar loop equation we
obtain
$$
\langle \hat w(z)\rangle|_{\rm planar} = -{2\over 3}
(t^{1/3}+2z)\sqrt{z-t^{1/3}}.\eqn\wplanar
$$
On the other hand, scaling the loop equation itself, we arrive at
$$
9(\langle \hat w(z)\rangle^2+\kappa^2 \langle \hat
w(z) \hat w(z)\rangle_c) = (24+6c)t - 12\langle P\rangle z+ 16z^3.
$$
With $c=-14/3$, $\langle P\rangle|_{\rm planar} = t^{2/3}$ we recover
precisely \wplanar. We have shown that operations of taking the
continuum limit and solving the loop equations commute.

Now we turn to $n=2$, $m=0$. According to \gscale\ $G(p)|_{\rm
univ.} = \langle \hat w(z)\rangle$ and universal loop equation should
appear at ${\cal O}(1)$ in scaled discrete equation. Indeed, for
$p_c^2=R_c=4$ we obtain
$$
\eqalign{0 &= \langle \hat w(z)\rangle^2 +\kappa^2 \langle \hat
w(z)\hat w(z)\rangle_c,\cr
0 &=\langle \hat w(z)\rangle|_{\rm planar}.\cr}\eqn\wscale
$$
For $n>2$, $m=0$ negative powers of $a$ in \gscale\ give immediately
$\langle \hat w(z)\rangle =0$.

\chapter{Appendix A}
In this appendix we examine with some detail potentials of the
form $V(\sum \alpha_r t_r) = \sum_{m} h_{m}(\alpha\cdot t)^{m}$.
These potentials already contain complicated couplings between
different traces, but they are still explicitly soluble and serve
as multicritical potentials for $m, n >1$ critical points.

The general matrix in \udef\ simplifies in this case to
$$
U'' = U \otimes \alpha
$$
where $\alpha, U$ denote the vectors of components $\alpha_k$ and
$$
U_k = V'' A_k \sum_{p}{p^{2}\over p+k} A_p \alpha_p
$$
(recall $A_k = \ap$). This structure makes the inversion of $1-U''$
trivial. Given that
$$
\partial_R x = D_R x + (U\cdot\alpha)\partial_R x =
D_R x_0 A + U(\alpha\cdot\partial_R x)\eqn\otra
$$
we may solve $\partial_R x_k$ as
$$
\partial_R x_k = D_R x_0 {(1-U\cdot\alpha)A_k +U_k \alpha
\cdot A\over 1-U\cdot\alpha}
$$
In particular
$$
\partial_R x_0 = {D_R x_0\over 1-U\cdot\alpha}(1-U\cdot\alpha+
U_0 \alpha\cdot A)
$$
and the polymer criticality condition reads:
$$
1-U\cdot\alpha+U_0 \alpha\cdot A =
1+V'' \sum_{p,k}{pk\over p+k}A_p A_k \alpha_p \alpha_p
\sim (R-R_c)^{n}\eqn\policrit
$$
Analogous considerations to those in \otra\ lead to
$$
\partial_{\beta} x_k = {24\over k+2}A_k\left({R\over 4}\right)^{2} +
U_k {\alpha\cdot D_{\beta} x\over 1-U\cdot\alpha}
$$
With these ingredients at hand we can compute $dx_k\over d\beta$,
whose scaling for large k determines the one-loop correlator,
$$
{dx_k\over d\beta}\sim -A_k {D_{\beta} x_0 (1-U\cdot\alpha)+
U_0 \alpha\cdot D_{\beta} x \over 1+W}
$$
for $1+W \equiv 1-U\cdot\alpha + U_0 \alpha\cdot A$. Assuming
an n-th order critical point for polymers in \policrit\ we find the
large k asymptotics
$$
{dx_k\over d\beta} \sim \ak (R-R_c)^{-n}
$$
the same scaling as for $V(x_l)$ potentials, and we conclude
 ${dw(\ell)\over dt}=0$ here as well, for $n>1$.

Now we come to the definition of $n>1, m \ge 1$ critical points.
First we should simplify the expressions for the moments $x_k$ by
solving $V'$ from the string equation:
$$
V'(\alpha\cdot x) = {1-\sum p A_p g_p \over \sum p A_p \alpha_p}
$$
and we get the functions
$$
x_k= X_k(R,g_p,\alpha_p)=\sum_{p}{p^{2}\over p+k} A_p A_k g_p +
{1-\sum p A_p g_p \over \sum p A_p \alpha_p}\sum_{p}{p^{2}\over
p+k} A_p A_k \alpha_p\eqn\momentos
$$
Notice that these functions $do$ depend on $g_2 =\beta$, so that
we still need the string equation to relate $\beta$ and $R$.

The Kazakov-like critical points come from
$$
RD_R x_0 ={\sum_{p,q}p^{2}q A_p A_q (g_p \alpha_q - \alpha_p g_q)
+ \sum_{p} p^{2} A_p \alpha_p \over \sum p A_p \alpha_p}
\sim (1-R)^{m}\eqn\kazakrit
$$
In this expression we assume $g_2 = \beta (R)$ solved from the string
equation.

Upon direct differentiation of \momentos\ we get
$$
{\partial X_k \over \partial R} = D_R x_0
{\sum_{p}{pk\over p+k} A_p A_k \alpha_p \over \sum p A_p \alpha_p}
$$
still proportional to $D_R x_0$, as in \dxkdr, but through a
non-trivial factor. If $R_c$ is the critical point in the polymer
phase then obviously
$$
{dX_k\over dR} = {\partial X_k\over \partial R}(R_c) +
{\partial X_k\over \partial \beta}{\partial \beta \over
\partial R}(R_c) = {\partial X_k\over \partial R}(R_c)\eqn\guapa
$$
Assume now a particular choice of $\alpha_p$ so that, in the vicinity
of $R_c = 1$ we have
$$
{\partial X_k\over \partial R}(R_c) \sim {(R_c -1)^{m}\over
\sum p A_p(R_c) \alpha_p} \sim (R_c -1)^{m-r}\eqn\casi
$$
for $0 < r \le m$. Differentiating now \policrit\ using \guapa\ and
\casi\ we arrive at:
$$
{dW\over dR}(R_c) \sim F(R_c)(1-R_c)^{2r} + G(R_c)V'''(\alpha\cdot x)
(1-R_c)^{m-r}
$$
with $F$ and $G$ regular functions at $R_c =1$. It is clear that
$dW/dR$ vanishes at $R_c =1$ for $0 < r < m$. If $r=m$ we need
additional tunning of $h_m$ parameters in the potential so that
$V'''(R_c = 1)=0$. Hence, we have explicitly shown how $n=2$
polymer critical points can coalesce with Kazakov critical points
of any order $m$. In general, the possibility of having ${dX_k
\over dR}\neq 0$ at the critical point rules out the problem for
arbitrary $n$.

In order to find an $n,m$ critical potential we are led to the
following hierarchy of tunnings: we first tune $\alpha_p$ as in
\casi. Then, for fixed $\alpha_p$, adjust $g_p$ to get \kazakrit\
and finally we may tune $h_m$ in \policrit. It is interesting to note
that the relative tunning among different traces in the non-linear
potential was essential to achieve the full spectrum of string
susceptibilities \suscrit\ quoted in the general discussion.

\chapter{Appendix B}

In this appendix we will calculate $\Phi(p,p)$ and $\Phi_4(p)$ needed in
the calculation of $1/N^2$-corrections in chapter 6. The identity
$$
0=\int {\cal D}\phi {d\over d\phi_{ab}} \left\{(e^{\ell\phi})_{ab}
\overn \tre{\ell_2}
\exp\left\{-N^2\left[\sum_{n=1}^s
g_n\left({1\over N} Tr \phi^{2n}\right)+V(t_4)\right]\right\} \right\}
$$
results in
$$
\eqalign{\sumones 2ng_n &\left({d\over d\ell}\right)^{2n-1} \leftl
\overn \tre{\ell_2} \overn \tre\ell\rightr +\cr
&\qquad +\sum_{m=1}^t 4m h_m\left({d\over d\ell}\right)^3
\leftl
\overn \tre{\ell_2} \left(\overn Tr \phi^4\right)^{m-1} \overn
\tre\ell\rightr =\cr
= \intol d\ell' &\leftl \overn \tre{\ell'} \overn \tre{(\ell-\ell')}
\overn \tre{\ell_2} \rightr + {\ell_2\over N}\leftl \overn
\tre{(\ell+\ell_2)} \rightr.\cr}
\eqn\newloop
$$
After expanding \newloop\ in terms of connected correlation functions,
using \aploop\ and going to the planar limit one is left with
$$
\eqalign{\sumones 2n\gtil n \left({d\over d\ell}\right)^{2n-1}
\Phi_{\ell,\ell_2} = 2\intol d\ell' &G_\ell \Phi_{\ell-\ell',\ell_2} +
\ell_2 G_{\ell+\ell_2} \cr
&-4V''(x_2) \left({d\over d\ell}\right)^3 G_\ell \leftl \tre{\ell_2}
Tr \phi^4\rightr_c.\cr}\eqn\nlpl
$$
The Laplace transform of \nlpl\ is
$$
\eqalign{2M(p) \sqrt{p^2-R} \Phi(p,p_2) = Q(p,p_2)&+ {\partial\over
\partial p_2}{G(p)-G(p_2)\over p-p_2} \cr
&-4V''(x_2) \Phi_4(p_2) (p^3 G(p)-p^2 -x_1),\cr}
\eqn\lapnl
$$
where
$$
Q(p,p_2) = \sum_{n=2}^s 2n \gtil n\sum_{i=1}^{2n-2} p^{2n-2-i}
\Phi_i(p_2).
$$
For our even matrix integral measure, $\Phi(p,p')$ splits into even
and odd parts.
\def\trtr#1#2{\langle Tr\phi^{#1} Tr\phi^{#2}\rangle_c}
$$
\eqalign{\Phi(p,p') &= \sum_{m,n=1}^\infty p^{-m-1} (p')^{-n-1} \trtr m
n,\cr
&=\sum_{m,n=1}^\infty  p^{-2m-1} (p')^{-2n-1} \trtr{{2m}}{{2n}} +\cr
&\quad +\sum_{m,n=1}^\infty p^{-2m} (p')^{-2n} \trtr{{2m-1}}{{2n-1}}
 = \Phi_{\rm even}(p,p')+\Phi_{\rm odd}(p,p').\cr}
$$
We will calculate $\Phi_{\rm even}$ using \dxk\ to obtain
$\trtr{{2j}}{{2k}} = -dx_k/dg_j$ for $l=2$. To get $\Phi_{\rm odd}$ one
can, following [\ambmak], expand \lapnl. The coefficient of $1/p_2^{2m}$
will determine $\Phi_{2m-1}(p)$. Since all explicit dependence on
$V''$ drops out of this expansion, $\Phi_{\rm odd}$ will have the same
form as in [\ambmak]:
$$
\Phi_{\rm odd} (p,p) ={2p^2 R-R^2\over 8p^2(p^2-R)^2}.
$$

As for the even part, in calculating
$$
\eqalign{\Phi_{\rm even}(p,p) &= -\sum_{k,j=1}^\infty {dx_k\over dg_j}
p^{-2k-2j-2},\cr
\Phi_4(p) &= -\sum_{k=1}^\infty {dx_k\over dg_2} p^{-2k-1},\cr}
$$
we will need the following sums:
$$
\eqalign{s_1(a) &= \sum_{k=1}^\infty {k\over k+2} \bco k \left({a\over
4}\right)^{k+2},\cr
s_2(a) &=\sum_{k,j=1}^\infty {jk\over j+k}\bco j\bco k \left({a\over
4}\right)^{k+j}.\cr}
$$
One can evaluate them as follows:
$$
\eqalign{s_1(a) &= \sum_{k=1}^\infty\bco k\left({a\over
4}\right)^{k+2} - \sum_{k=1}^\infty {2\over k+2}\bco k \left({a\over
4}\right)^{k+2} ,\cr
&= \left({a\over
4}\right)^{2} \left[{1\over\sqrt{1-a}}-1\right] - {1\over 8}
\int\limits_0^a dt t\left[{1\over\sqrt{1-t}}-1\right],\cr
&= {8-4a-a^2\over 48\sqrt{1-a}} - {1\over 6},\cr
s_2(a) &= \int\limits_0^a dt {1\over t} \left(\sum_{k=1}^\infty k
\bco k\left({t\over 4}\right)^{k} \right)^2,\cr
&= \int\limits_0^a dt {1\over t} \left(t {\partial\over \partial t}
\sum_{k=0}^\infty \bco k\left({t\over 4}\right)^{k} \right)^2,\cr
&= {1\over 4}\int\limits_0^a dt t (1-t)^{-3} = {1\over 8}
\left({a\over 1-a}\right)^2.\cr}
$$
Now it is easy to show that
$$
\eqalign{\Phi_4(p) &={1\over 1+W}\left[{8p^4-4Rp^2-R^2\over
4\sqrt{p^2-R}} - 2p^3\right],\cr
\Phi(p,p) &= {R\over 4(p^2-R)^2}\cr
&\; -{4W\over 9R^4(1+W)}\left[
128p^6+ {16p^3 (R^2+4Rp^2-8p^4)\over \sqrt{p^2-R}} + {R^3(8p^2+R)\over
p^2-R}\right].\cr}
$$

\endpage
\refout
\end